\documentclass[twocolumn,showpacs,preprintnumbers,amsmath,amssymb]{revtex4}
\input psfig.sty
\input colordvi.sty

\usepackage{graphicx}
\usepackage{dcolumn}
\usepackage{bm}

\def\be{\begin{equation}}
\def\ee{\end{equation}}

\def\apj{{\it Astrophys.~J.}}

\begin{document}


\title{Testing dark energy beyond the cosmological constant \emph{barrier}}

\author{J. S. Alcaniz} \email{alcaniz@dfte.ufrn.br}

\affiliation{Departamento de Fisica, Universidade Federal do Rio
Grande do Norte, C.P. 1641, Natal - RN, 59072-970, Brasil }

\vspace{0.2cm}

\date{\today}

\begin{abstract}

Although well motivated from theoretical arguments, the cosmological constant 
\emph{barrier},
i.e., the imposition that the equation-of-state parameter of dark energy ($\omega_x
\equiv p_x/\rho_x$) is $\geq -1$, seems to introduce bias in the parameter
determination from statistical analyses of observational data. In this regard,
\emph{phantom} dark energy or \emph{superquintessence} has been proposed in which the
usual imposition $\omega \geq -1$ is relaxed. Here, we study possible observational
limits to the \emph{phantom} behavior of the dark energy from recent distance estimates
of galaxy clusters obtained from interferometric measurements of the Sunyaev-Zel'dovich
effect/X-ray observations, Type Ia supernova data and CMB measurements. We find that
there is much
\emph{observationally} acceptable parameter space beyond the $\Lambda$ \emph{barrier},
thus opening the possibility of existence of more exotic forms of energy in the
Universe.

\end{abstract}

\pacs{98.80.Es; 95.35.+d; 98.65.Cw}
\maketitle

\section{Introduction}

Dark energy or \emph{quintessence} is the invisible fuel that seems to drive the curent
acceleration of the Universe. Phenomenologically, this energy component is usually
described
by an equation-of-state parameter $\omega_{x}$ which represents the ratio of the dark
energy pressure to its energy density, $\omega_x \equiv p_x/\rho_x$. In order to achieve
cosmic  acceleration, Einstein Field Equations (EFE) require $\omega_x$ to be  $< - 1/3$
for a universe
described by a single component whereas for a dark matter/energy dominated universe the
required value, $\omega_x < - (\Omega_m/3\Omega_{x} + 1/3)$, depends on the ratio
between the baryonic/dark matter ($\Omega_m$) and dark energy density parameters
($\Omega_{x}$). In other words, what EFE mean with these upper limits is that any
physical field with positive energy density and negative
pressure,
which violates the strong energy condition ($\rho + 3p > 0$), may cause antigravity
regimes (see \cite{carrol} for a review on classical energy conditions).

Since cosmic acceleration from EFE provides only an upper limit to
$\omega_x$, a point of fundamental importance associated with this parametrization for
the dark energy
equation of state (ES) is related to the physical and/or observational \emph{lower}
limits that
may
be imposed on the parameter $\omega_x$. Physically, if one wants dark energy to be
stable, then it must obey the null energy condition which, in the
Friedmann-Robertson-Walker metric, is equivalent to $\rho + p > 0$. This energy
condition implies $\omega_x \geq -1$ when applied to a dark energy component described
by
$\omega_x \equiv p_x/\rho_x$ or, equivalently, that the vacuum energy density or
the cosmological
constant($\Lambda$), which is characterized by $\omega_x = -1$, would constitute
the natural lower
limiting case. Following this reasoning, firstly
explicited in \cite{garn}, a number of
theoretical and observational
analyses in which the restriction $-1 \leq \omega_x < 0$ is imposed  have appeared in
the
recent literature \cite{xmatter}. However, by focussing our attention only on the
observational
side, what would current
observations have to tell us about that? As well observed by Caldwell \cite{cald}, it
is curious that most of the observational constraints on $\omega_x$ are consistent with
models that go right up to the $\omega_x = -1$ border. Thus, paraphrasing him, one might
ask what lies on the other side of the cosmological constant \emph{barrier}.

The answer to this question has been given by several authors who also have
pointed out some strange properties of \emph{phantom} dark energy ($\omega_x < -1$) as,
for instance, the fact that its energy density increases with the expansion of the
Universe in contrast with usual \emph{quintessence} ($\omega_x \geq -1$); the
possibility
of rip-off of the large and small scale structure of matter; a possible occurence of
future curvature singularity, etc. \cite{cald1}. Although having these unusual
characteristics, a \emph{phantom} behavior is predicted by several scenarios, e.g.,
kinetically driven models \cite{chiba1} and some versions of brane world
cosmologies \cite{sahni} (see also \cite{carrol} and references therein). Moreover,
from the
observational point of view, \emph{phantom} energy is found to be compatible with most
of the classical cosmological tests and seems to provide a better fit to type Ia
supernovae (SNe Ia) observations than do $\Lambda$CDM or generic \emph{quintesence}
scenarios ($\omega_x \geq  -1$) \cite{pad}. Therefore, given our state of complete
ignorance about the nature of dark energy, it is worth asking whether current
observations are able to shed some light on the other side of the $\Lambda$
\emph{barrier}.

\begin{figure*}
\centerline{\psfig{figure=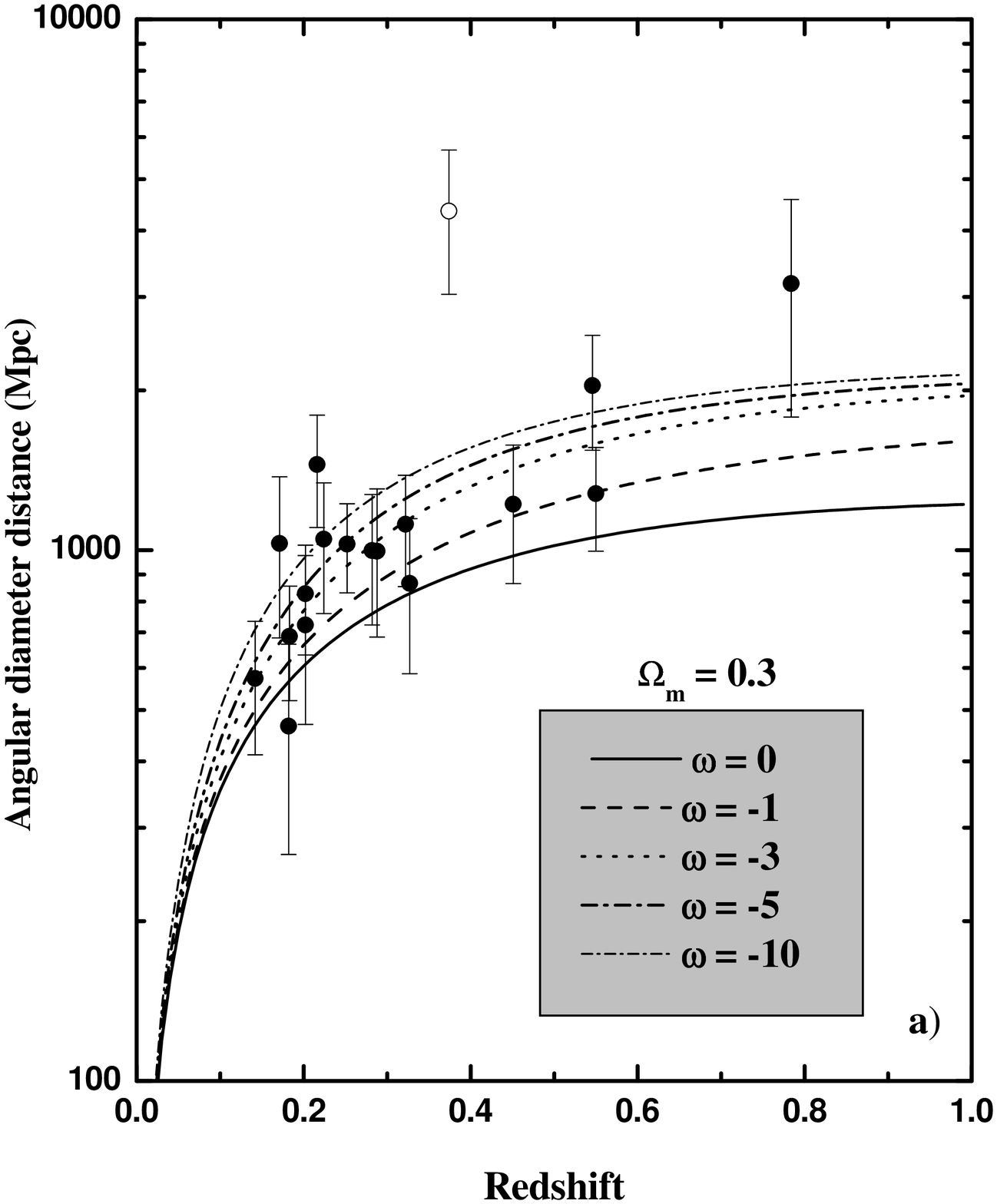,width=2.4truein,height=3.7truein,angle=0}
\psfig{figure=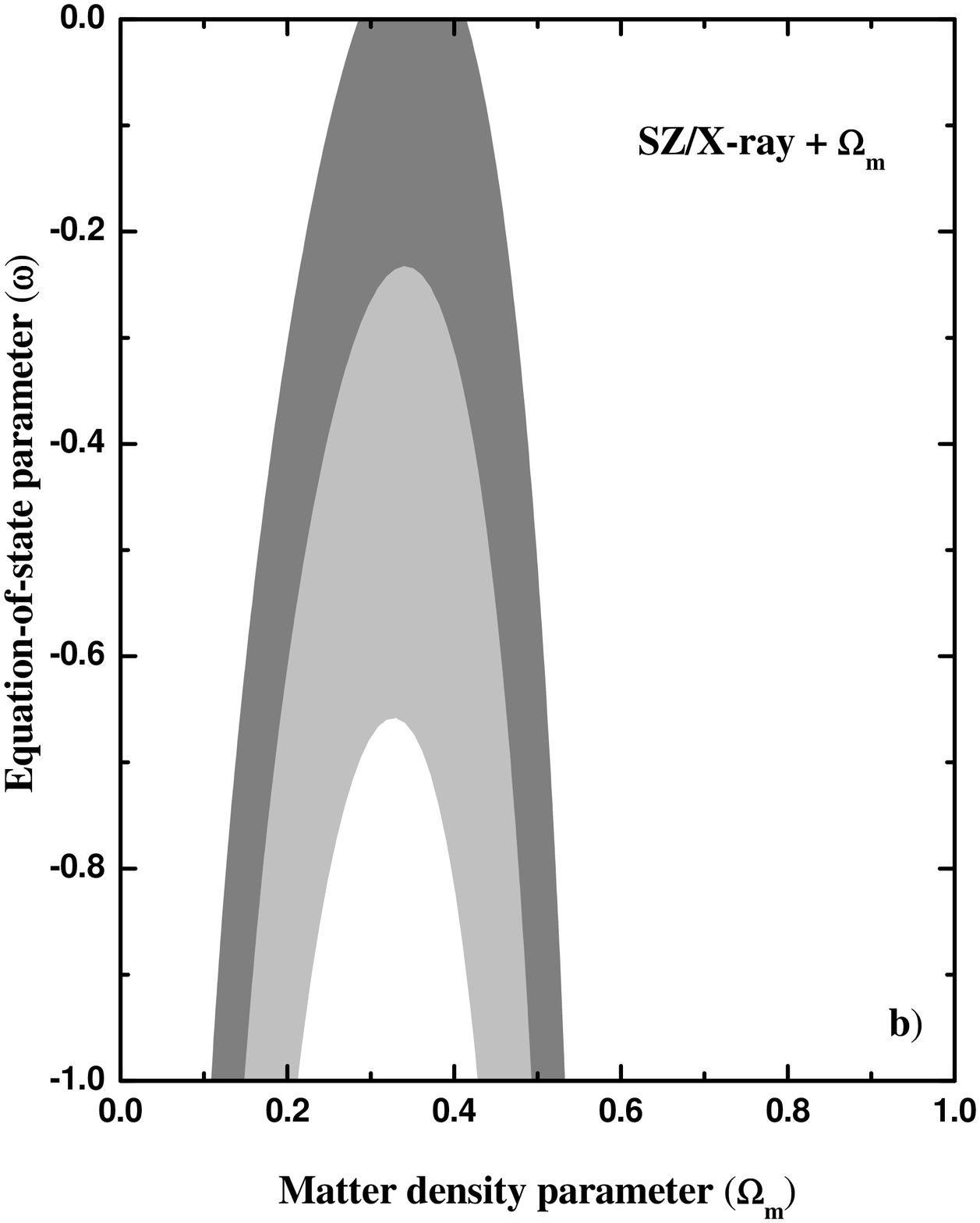,width=2.4truein,height=3.7truein,angle=0}
\psfig{figure=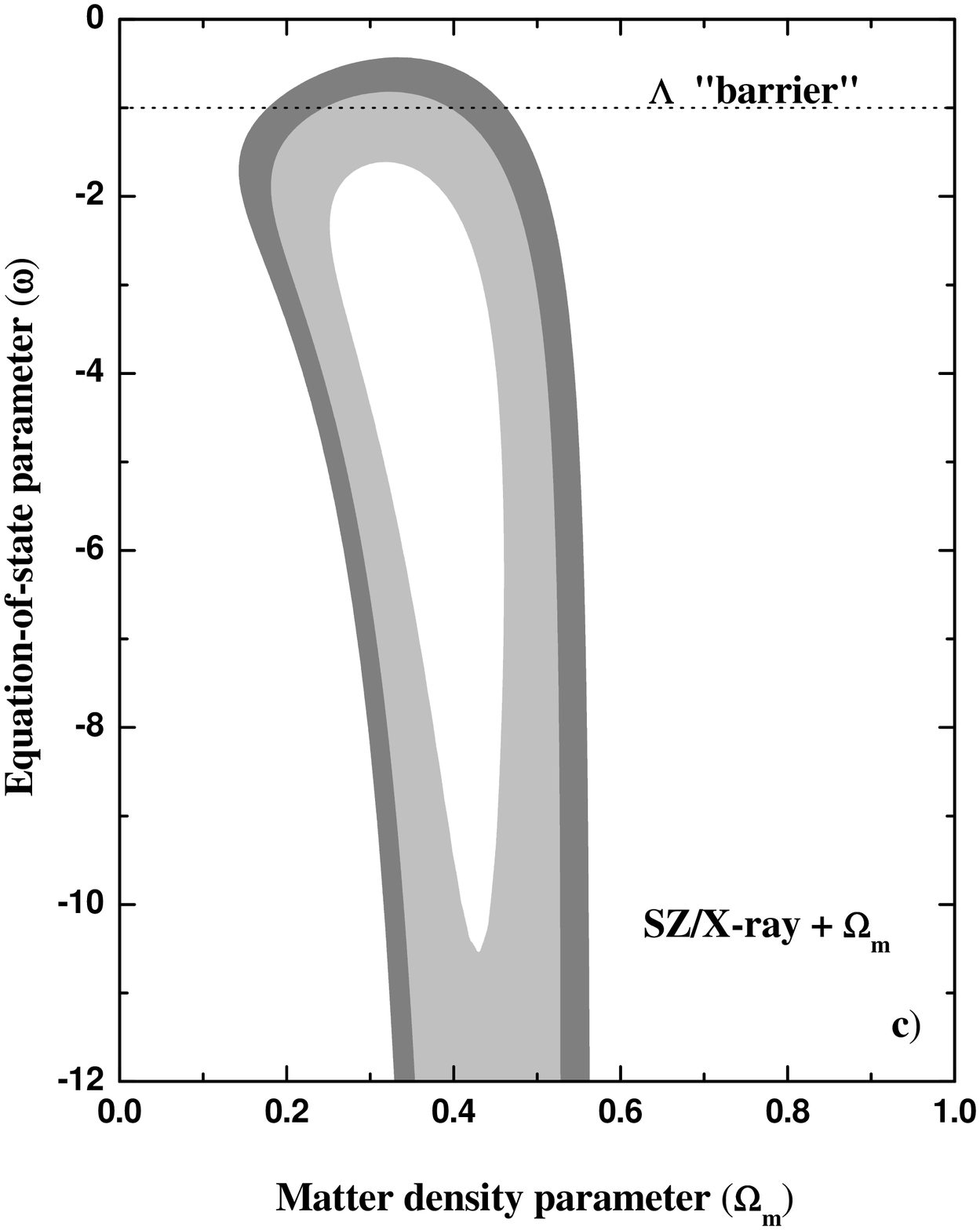,width=2.4truein,height=3.7truein,angle=0}
\hskip 0.1in}
\caption{{\bf{a)}} SZE/X-ray determined distances for 18 clusters as a function of
redshift for a fixed value of $\Omega_m = 0.3$ and selected values of the ES parameter.
The open circle corresponds to Abell 370 cluster which has been excluded from the
statistical analyses. {\bf{b)}} Confidence regions (68$\%, 95$\% and 99\%) in the
$\Omega_{m} - \omega$ plane provided by the SZE/X-ray ADD data from Reese {\it et al.}
\cite{reese} by assuming a Gaussian prior on the matter density parameter
$\Omega_{\rm{m}} = 0.35 \pm 0.07$. {\bf{c)}} The same as in Panel 1b with the $\Lambda$
\emph{barrier} removed.}
\end{figure*}

Our aim, in this \emph{Letter}, is to seek possible observational limits to the
\emph{phantom} behavior of the dark energy ES, as well as to detect the bias in the
ES parameter determination due to the imposition $\omega_x \geq -1$, from recent
distance estimates of
galaxy clusters obtained from interferometric
measurements of the Sunyaev-Zel'dovich effect (SZE) and X-ray observations. We
use, for that, the largest homogeneously analyzed sample of the SZE/X-ray clusters with
angular
diameter distance (ADD) determinations thus far, as provided by Reese {\it et al.}
\cite{reese}. In order to constrain more precisely regions of the parameter space, we
also combine SZE/X-ray ADD data with the newest SNe Ia sample of the \emph{Supernova
Cosmology Project} \cite{knop}, recent determinations of the matter density parameter,
WMAP distance estimates \cite{wmap} and the latest measurements of the Hubble parameter
as given by the \emph{HST} key
project \cite{freedman}. In agreement with other independent analyses, it is shown that
with or without such a combination, these observational data
do prefer the supernegative behavior of the dark energy equation of state.

\section{SZE, X-ray Emission and Distance Estimates}

Among the sources of temperature
flutuations in the Cosmic Microwave Background Radiation (CMBR), a small distortion due
to inverse Compton scattering of CMBR photons passing through an intracluster medium is
of particular importance to estimating distances to galaxy clusters. This is so because
for a given temperature such effect, known as Sunyaev-Zel'dovich effect \cite{zeld}, is
proportional to the line integral of the eletron number density through the cluster,
$\Delta T \propto \int{n_e T_e d\ell}$, while the X-ray bremsstrahlung surface
brightness scales as $S_X \propto \int{n_e^{2} d\ell}$. Thus, by using X-ray
spectroscopy to find the temperature of the gas and by making some assumptions on the
cluster geometry, the distance to the cluster may be estimated (see \cite{birk} for
recent summaries).

By applying this technique, suggested long ago \cite{szxray}, Reese {\it et al.}
\cite{reese} determined the distance to 18 galaxy clusters with redshifts ranging from
0.14 to 0.78, which constitutes the largest homogeneously analyzed sample of the
SZE/X-ray clusters with distance determinations thus far. From these intermediary and
high-$z$ measurements, the authors estimated the Hubble parameter for three different
cosmologies, with the uncertainties agreeding with the \emph{HST} key project results
\cite{freedman}, which probes the expansion rate in the nearby universe. Since the
redshift range of the
galaxy cluster sample is comparable to the intermediary and high-$z$ SNe Ia data
compiled by the \emph{Supernova Cosmology Project} \cite{knop,perlmutter} and the
\emph{High-$z$ Supernova Team} \cite{riess}, we understand that it may also provide an
independent crosscheck of the cosmic acceleration mechanism. Thus, in what folows, we
use these data as well as a combination of them with SNe Ia measurements to place
observational limits on the ES parameter of the \emph{phantom} dark energy.

\begin{figure*}
\centerline{\psfig{figure=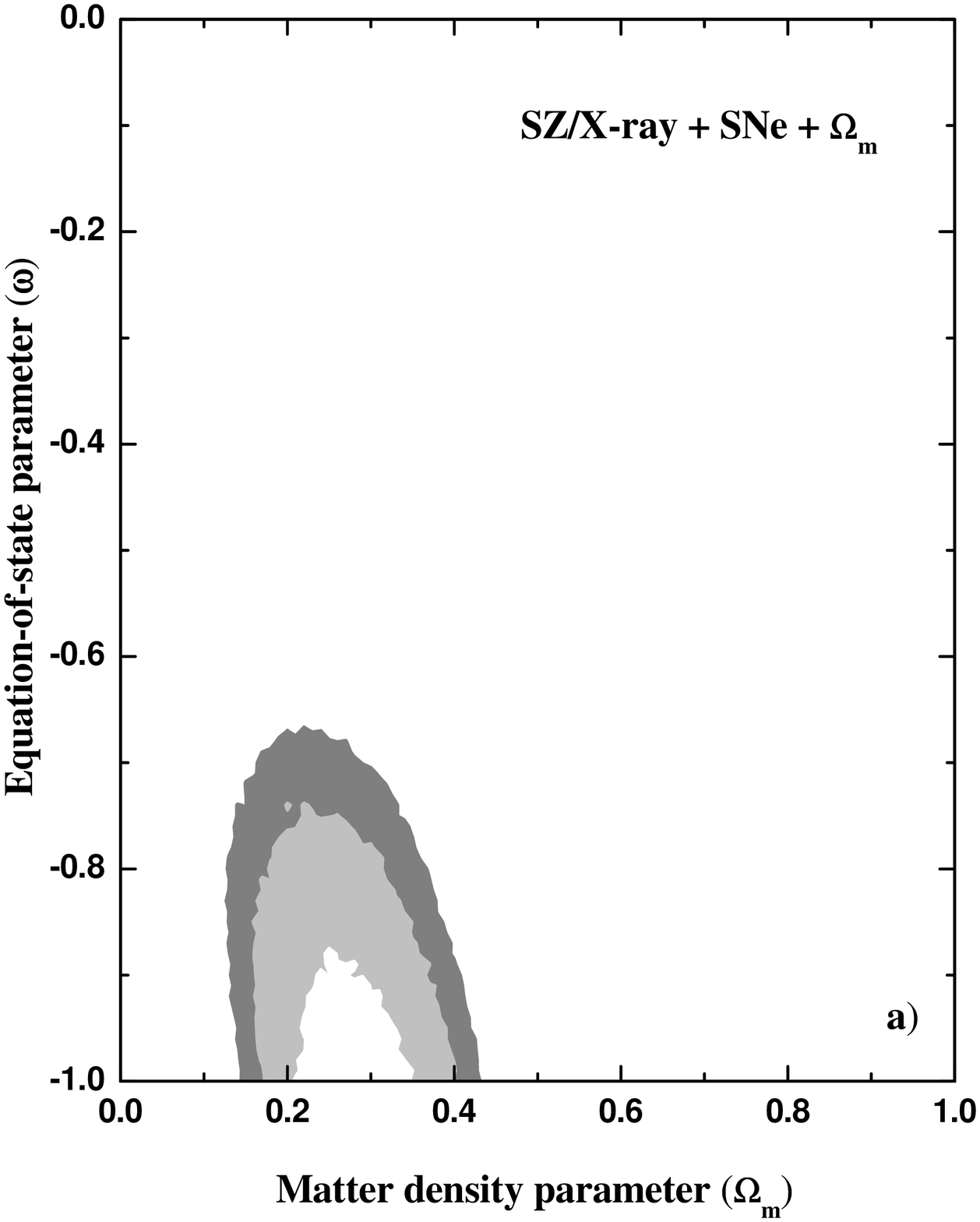,width=2.4truein,height=3.7truein,angle=0}
\psfig{figure=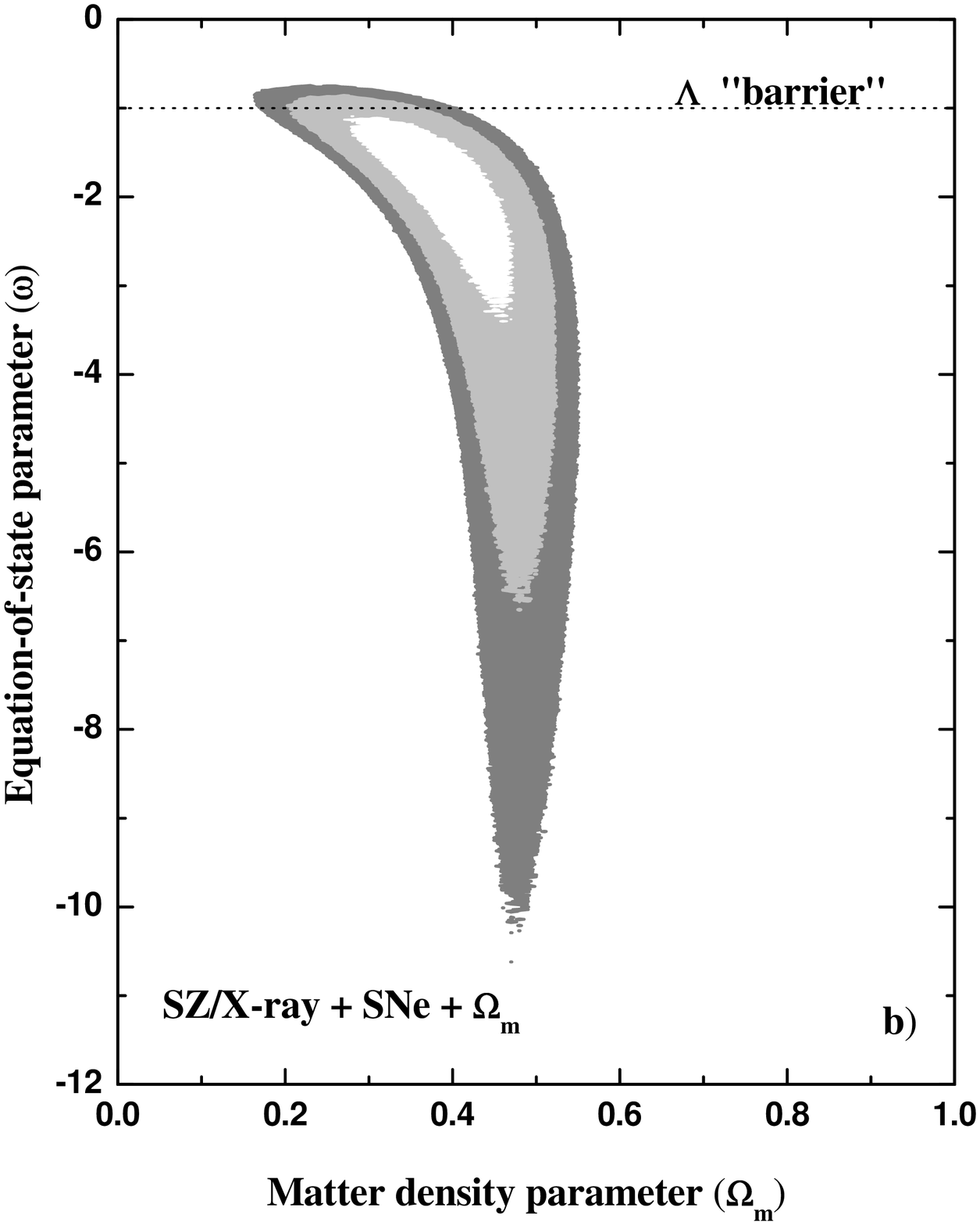,width=2.4truein,height=3.7truein,angle=0}
\psfig{figure=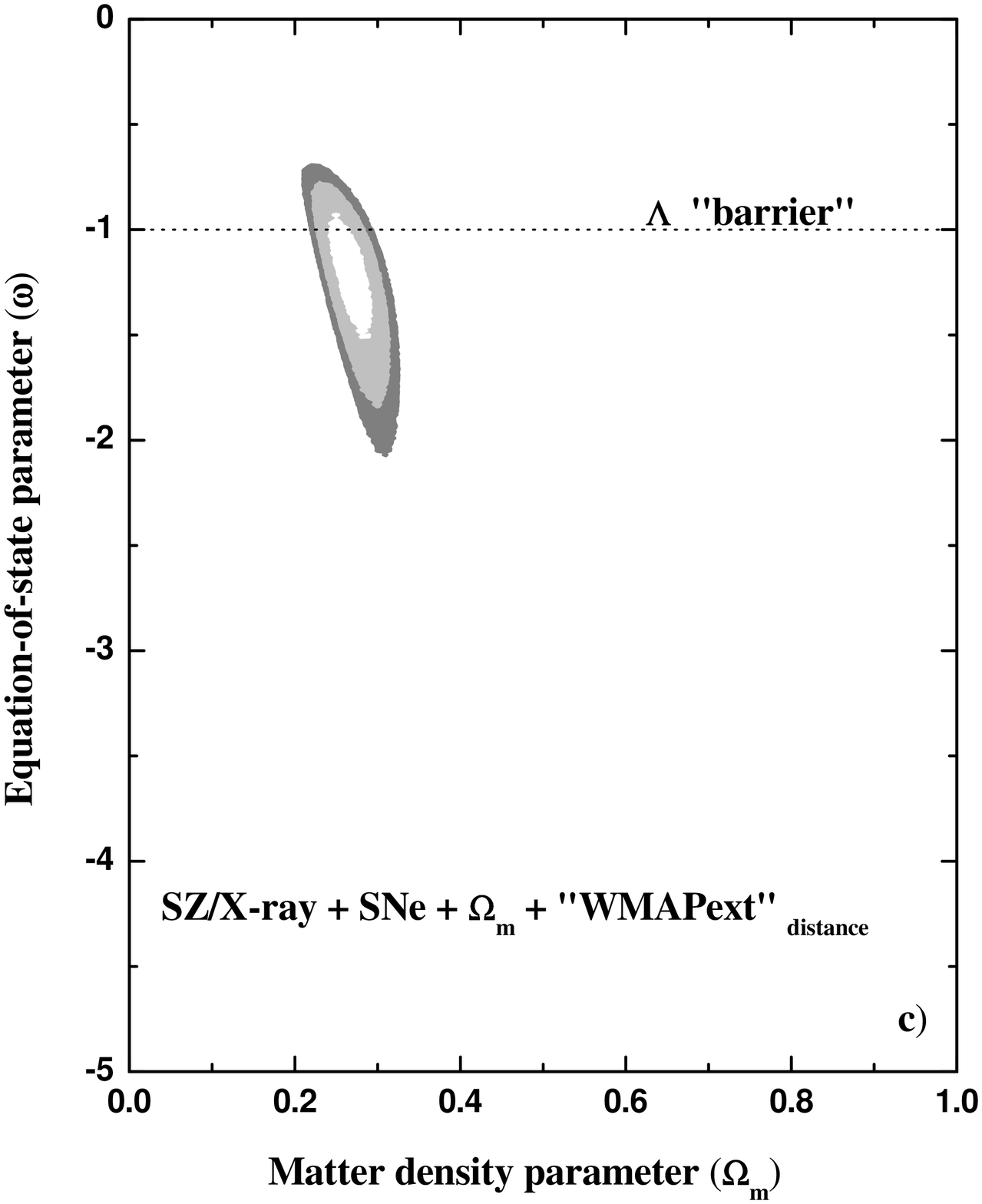,width=2.4truein,height=3.7truein,angle=0}
\hskip 0.1in}
\caption{{\bf{a)}} The likelihood contours in the $\Omega_m - \omega$ plane for the
joint SZE/X-ray ADD + $\Omega_m$ + SNe Ia analysis
described in the text. The contours correspond to 68\%, 95\% and
99\% confidence levels. {\bf{b)}} The same as in Panel 2a with the $\Lambda$
\emph{barrier} removed. For this analysis the best fit values
are located at $\omega = -1.7$ and $\Omega_{m} = 0.38$. {\bf{c)}} The same as in Panel
2b with the ``WMAPext'' constraint on the angular size distance to the decoulping
surface at $z = 1089$. The best-fit model converges to $\omega = -1.2$ (and $\Omega_m =
0.27$), with a 68\% confidence bound of $-1.38 \leq \omega \leq -1.09$.}
\end{figure*}

\section{Analysis}

With the usual assumption that the effective equation of state, $\omega
\sim
\int{\omega_{x}(z)\Omega_x(z)dz}/\int{\Omega_x(z)dz}$, is a good approximation for a
wide class of dark energy scenarios \cite{stein}, the angular diameter distance as a
function of the redshift can be written as
\begin{equation}
{\cal{D}}_A(z;\Omega_m,\omega) = \frac{3000h^{-1}}{(1 +
z)}\int_{o}^{z}\frac{dz'}{{\cal{E}}(z';\Omega_m,\omega)} \quad \mbox{Mpc},
\end{equation}
where the dimensionless function ${\cal{E}}(z';\Omega_m, \omega)$ is given by
\begin{equation}
{\cal{E}} = \left[\Omega_m(1 + z')^{3} +
(1 -\Omega_m)(1 + z')^{3(1 + \omega)}\right]^{1/2}.
\end{equation}

Figure 1a shows the SZE/X-ray determined distances for 18 clusters as a function of
redshift for a fixed value of $\Omega_m = 0.3$ and selected values of the ES parameter.
Note that Abell 370 cluster (the open circle) is clearly an outlier in the sample so
that, following \cite{reese,zhu}, we exclude it from the statistical analyses that
follow.
In
Fig. 1b we show the confidence regions (68\%, 95\% and 99\%) in the plane $\Omega_m -
\omega$ from SZE/X-ray ADD data. Since we have nowadays good estimates of the dark
matter density \cite{calb}, we have assumed a Gaussian prior on
the matter density parameter, i.e., $\Omega_{\rm{m}} = 0.35 \pm 0.07$.
Such a value, which is in good agreement with dynamical estimates on scales up to about
2$h^{-1}$ Mpc \cite{calb}, is derived by combining the ratio of baryons to
the total mass in clusters determined from SZE/X-ray measurements with the latest
estimates of the
baryon density $\Omega_{\rm{b}} = (0.020 \pm 0.002)h^{-2}$
\cite{burles} and the final value of the Hubble parameter obtained
by the {\it {HST}} key Project $H_o = 72 \pm 8$
${\rm{km.s^{-1}.Mpc^{-1}}}$ \cite{freedman}. As the Figure shows, given the $\Lambda$
\emph{barrier}, the best-fit converges to $\omega = -1$ (and $\Omega_m = 0.32$), with a
68\% confidence bound of $\omega \leq -0.84$. 

A generalization of this analysis to a parameter space that extends $\omega$ to values
smaller than $-1$ is presented in Figure 1c (in all extended analyses, we have used a
$\chi^{2}$ minimization for the range of $\Omega_m$ and $\omega$ spanning the interval
[0,1] and [-15,0], respectively). There, it is shown that there is much
\emph{observationally} acceptable parameter space beyond the $\Lambda$ \emph{barrier},
in fully
agreement with other similar analyses \cite{hann,sing,johri}. In actual fact, the
best-fit model
for these data sets occurs for $\Omega_m = 0.36$ and $\omega = -3.5$ ($\chi^{2}_{min} =
10.8$) with a
68\% confidence bound of $-5.5 \leq \omega \leq -2.2$ ($0.30
\leq \Omega_m \leq 0.41$). In particular, this best-fit model corresponds to a
very accelerating universe with deceleration parameter $q_o \simeq -2.8$ and total
expanding
age of $9.7h^{-1}$ Gyr. It is also worth noticing that extreme values of
$\omega$ are allowed because for intermediary and high redshifts, angular diameter
distances become quite insensitive to large variations of the ES parameter (see Fig.
1a). For example, at $z = 0.78$ (the redshift of MS1137, the farthest galaxy cluster),
the angular diameter distance for $\Omega_m = 0.3$ and $\omega = -3$ (${\cal{D}}_A
\simeq 1840$ Mpc) is only $\sim 10\%$ smaller than in a model with the same amount of
dark
matter and $\omega = -10$ (${\cal{D}}_A \simeq 2045$ Mpc). This particular behavior is
quite similar to what happens in analyses involving age estimates. There,
like here, the function of the cosmological parameters ($\Omega_m$ and $\omega$)
quickly asymptotes for large values of $\omega$ \cite{krauss}.

In our search for possible lower limits to the ES of the \emph{phantom} component, we
now perform a joint analysis of SZE/X-ray ADD and SNe Ia data. For that, we use the
newest SNe Ia sample of the \emph{Supernova Cosmology Project} \cite{knop} (with strech
and extinction correction applied) and follow the analytical marginalization method for
the ``zero point magnitude'' ${\cal{M}}$ as given in Ref. \cite{gol}. The results of
the present analysis
are shown in Panels 2a and 2b. In Panel 2a we show the 68\%, 95\% and 99\% c.l. in the
$\Omega_m - \omega$ plane by imposing the $\Lambda$ \emph{barrier} ($\omega \geq -1$).
From this combination of observational data sets we find that the best-fit model occurs
exactly on the $\omega = -1$ border with $\Omega_m = 0.29$ and $\chi^{2}_{min}/\nu
\simeq 1.26$. At 95\% c.l., we obtain $\omega \leq -0.83$ and $0.19 \leq \Omega_m \leq
0.37$. Figure 2b generalizes the previous analysis to include more negative values of
the
dark energy ES. Again, we find that there is much acceptable parameter space beyond the
line $\omega = -1$ and that the confidence regions are modified by its presence, what
clearly indicates the existence of bias in the parameter
determination due to the $\Lambda$ \emph{barrier}. This particular analysis provides a
68\% confidence bound of $-1.98 \leq \omega \leq -1.42$ and $0.30 \leq \Omega_m \leq
0.45$, with the best-fit model happening at $\omega = -1.7$ and $\Omega_m = 0.38$
($\chi^{2}_{min}/\nu \simeq 1.2$), which corresponds to a accelerating universe with
deceleration parameter $q_o \simeq -1.0$ and total expanding age of $9.3h^{-1}$ Gyr.
If one combines this 68\% confidence
bound on $\omega$ with the upper limit from EFE, one would have $-1.9 \leq \omega <
-1/3$ instead of the usual $-1 \leq \omega < -1/3$. 

At this point, it is important to observe that the very
low-$\omega$ region of the above analyses can be considerably reduced by
combining them with high-$z$ data as, for instance, the current CMB
measurements (see, e.g., \cite{melc}). To better visualize that, Fig. 2c shows the
results of a combined test involving SZE/X-ray ADD + SNe Ia data and the
``WMAPext'' constraint (which includes other CMB experiments in addition to WMAP) on
the angular size distance to the decoulping surface at $z = 1089$, i.e., $d =
14.0^{+0.2}_{-0.3}$ Gpc \cite{wmap}. This analysis shows that the best-fit model moves
up to converge at $\omega = -1.2$ (and $\Omega_m = 0.27$), with a
68\% confidence bound of $-1.38 \leq \omega \leq -1.09$. These results, along with the
gradual decrease of the low-$\omega$ region seen from
Figs. 1a to 2c, clearly show that SNe Ia and CMB measurements dominate the analyses
over SZE/X-ray ADD data, which can be directly associated with the current systematics
uncertainties on these latter measurements. As commented in Ref. \cite{reese}, such
systematics are observationally approachable and will be addressed in the coming years
through the current generation of X-ray satellites (Chandra \& XMM-Newton) and radio
observatories (OVRO, BIMA \& VLA). Surely, these improvements will be very welcome once 
SZE/X-ray determined distances are measurements independent of the extragalactic
distance ladder that may provide distance to high-$z$ galaxy clusters. With such a
future sample of high-$z$ objects, it is expected that SZE/X-ray ADD data will be able
to provide a valuable independent check of SNe Ia and primary CMB power spectrum
results. 

We now compare our results with other recent
limits on the ES parameter of the \emph{phantom} energy derived from independent
methods. For
example, in Ref. \cite{hann} data from CMBR, large scale structure (LSS) and SNe Ia were
combined to find a 95\% confidence bound of $-2.68 < \omega < -0.78$. Such results
agree with the constraints obtained from a combination of Chandra observations of the
X-ray luminosity of galaxy clusters with independent measurements of the baryonic
matter density and the latest measurements of the Hubble parameter. From this latter
analysis, it was found $-2.0 \leq \omega \leq -0.6$ at 68\% c.l. \cite{alcaniz} while a
combination of these X-ray data with measurements of the angular size of milliarcsecond
radio sources provide $-2.22 \leq \omega \leq -0.62$ at 95\% c.l. \cite{zhu1}.
Recently, constraints from several CMBR experiments (including the latest WMAP results)
along with LSS data, Hubble parameter measurements from the \emph{HST} key
project and SNe Ia data were obtained, with the ES parameter ranging from $-1.38$ to
$-0.82$ at 95\% c.l. \cite{melc}. More recently, the authors of Ref. \cite{sing} used a
sample of 57
SNe Ia to find a 95\% confidence bound of $-2.4 < \omega < -1$ whereas estimates of the
age of the Universe as given by WMAP ($t_o = 13.7 \pm 0.2$ Gyr) provide $-1.18 < \omega
< -0.93$, which corresponds to an accelerating scenario with the deceleration parameter
$q_o$ lying in the range $-0.8 < q_o < -0.52$ \cite{johri}. All these results agree at
some level with the ones found in this work.

\section{Conclusion}

This paper, as many of its predecessors, is mainly motivated by our 
present state of ignorance concerning the nature of the so-called dark energy (or dark
pressure). In a first moment, vacuum energy density or cosmological constant was thought
of (also motivated by the \emph{old} age of the Universe problem) as the most viable
explanation for the evidence of cosmic acceleration as given by SNe
Ia observations. Observationally, $\Lambda$ remains as a good
candidate for dark
energy although, from a theoretical viewpoint, one has to face a fine-tuning of 120
orders
of magnitude in order to
make its \emph{observed} value compatible with quantum field theory expectations
\cite{lambdaP}. Later
on, a first generalization of this former description, in which a
``X-matter'' component with ES parameter ranging from a cosmological constant ($\omega =
-1$) to presureless matter ($\omega = 0$), was proposed as a possible
description for current observations \cite{xmatter}. More recently, a new
generalization, the
so-called \emph{phantom} energy, in which the $\Lambda$ \emph{barrier} ($\omega = -1$)
is removed, has received increasing attention among theorists. Naturally, all these
theoretical attempts to describe dark energy would not be valid without observational
support. But
that is not the case once several observational analyses support these parametrizations
for dark energy. Here, we have explored the prospects for constraining the
\emph{phantom} 
behavior of the dark energy from SZE/X-ray distance estimates of galaxy clusters, 
SNe Ia data and CMB-based distance estimates. We have shown that these data allow much
acceptable parameter space beyond
the line $\omega = -1$, what indicates not only the possibility of bias in the parameter
determination when the $\Lambda$ \emph{barrier} is imposed but also the possibility of
existence of more exotic forms of energy in the Universe. Naturally, we do not expect
such results to be completely free of observational and/or theoretical uncertainties,
mainly because there still exist considerable systematics uncertainties associated with
SZE/X-ray distance determinations.
What we do expect is that in the near future new sets of observations along with more
theoretical effort will be able to decide on which side of the $\Lambda$ {\it barrier}
lies the so far mysterious dark energy.

\begin{acknowledgments}
The author is very grateful to Z.-H. Zhu and F. A. da Costa for valuable discussions and
a critical
reading of the manuscript. 
This work is supported by the Conselho Nacional de
Desenvolvimento Cient\'{\i}fico e Tecnol\'{o}gico (CNPq - Brasil) and CNPq
(62.0053/01-1-PADCT III/Milenio).
\end{acknowledgments}


\end{document}